\documentclass[prd,floatfix,preprintnumbers]{revtex4}

\usepackage{graphicx}
\usepackage{graphicx,epsfig}
\usepackage{bm}
\usepackage{latexsym,amssymb,amsmath,float}

\newcommand {\ga} {\ {\raise-.5ex\hbox{$\buildrel>\over\sim$}}\ }
\newcommand {\la} {\ {\raise-.5ex\hbox{$\buildrel<\over\sim$}}\ }

\def\be{\begin{equation}}
\def\ee{\end{equation}}
\def\ba{\begin{eqnarray}}
\def\ea{\end{eqnarray}}

\begin{document}

\title{The cosmological evolution of ultralight axionlike scalar fields}
\author{Cameron E. Norton and Robert J. Scherrer}
\affiliation{Department of Physics and Astronomy, Vanderbilt University,
Nashville, TN  ~~37235}

\begin{abstract}
We examine the cosmological evolution of ultralight axionlike (ULA) scalar fields with potentials of
the form $V(\phi) = m^2 f^2[1 - \cos(\phi/f)]^n$, with particular emphasis
on the deviation in their behavior from the corresponding small$-\phi$ power-law approximations to these potentials:
$V(\phi) \propto \phi^{2n}$.  We show that in the slow-roll regime, when
$\dot \phi^2/2 \ll V(\phi)$, the full ULA potentials yield a more interesting range
of possibilities for quintessence than do the corresponding power law approximations.
For rapidly oscillating scalar fields, we derive the equation of state parameter
and oscillation frequency for the ULA potentials and show how they deviate
from the corresponding power-law values.  We derive an analytic expression
for the equation of state parameter that better approximates the ULA value than
does the pure power-law approximation.
\end{abstract}

\maketitle

\section{Introduction}

Scalar fields are ubiquitous in cosmology, serving as the
primary component of models for inflation \cite{Lyth,Allahverdi} and as ``quintessence"
models for the current accelerated expansion
of the universe
\cite{RatraPeebles,Wetterich,Ferreira,CLW,CaldwellDaveSteinhardt,Liddle,SteinhardtWangZlatev,
Copeland1}.
More recently, the possibility that a scalar field might
contribute transiently to the energy density has been proposed as a possible solution
to the ``Hubble tension," the discrepancy between direct local measurements of the Hubble parameter
and the value inferred from measurements of the cosmic microwave background (CMB) \cite{Freedman,Knox}.
In these scalar field solutions to the Hubble tension, the universe is never dominated by the energy density of the scalar field;
instead, the scalar field density reaches roughly 10\% of the total density in the universe
and then decays away \cite{Poulin1,Poulin2,Agrawal,Lin,Smith,Murgia}.

A particularly good fit to the Hubble data can be obtained
for a scalar field potential of the form
\begin{equation}
\label{ula}
V(\phi) = m^2 f^2[1-\cos(\phi/f)]^n.
\end{equation}
The case where $n=1$ is the well-studied axion potential, and following Ref. \cite{Poulin1},
we will refer to scalar fields with potentials given by Eq. (\ref{ula}) as ultralight axionlike
(ULA) fields.  Such potentials (with $n=1$)
were among the first proposed for quintessence \cite{Frieman} and have been extensively studied in that
context.  Larger values of $n$ might arise from higher-order instanton corrections \cite{Kappl}.  Other
cosmological consequences of scalar fields with potentials given by Eq. (\ref{ula}) are examined
in Refs. \cite{Poulin1,Poulin2,Capparelli}.

Most of these studies simply approximate the potential in Eq. (\ref{ula}) by the corresponding power
law appropriate for $\phi \ll f$, i.e., 
\begin{equation}
\label{powerula}
V(\phi) = \frac{1}{2^n} m^2 f^2 (\phi/f)^{2n}.
\end{equation}
This is
an excellent approximation in the limit of small $\phi$, but as shown in Ref. \cite{Smith}, the use
of this approximation can cause the evolution
of the scalar field to deviate significantly from the evolution for the full potential of Eq. (\ref{ula}).
For this reason, we undertake here a general study of scalar field evolution with ULA potentials, with
the goal of understanding the ways in which this evolution deviates from the corresponding power-law evolution.
Because these fields have been applied as both models for quintessence and as solutions of the Hubble tension,
we will
keep our discussion as general as possible, considering both the slow-roll and oscillatory phases
of the scalar field evolution.

In the next section, we examine the evolution
of ULA scalar fields in detail.  We first investigate
the initial slow-roll thawing evolution of these models and then their behavior when they oscillate rapidly.  We highlight the differences between the evolution of the ULA
models and the corresponding power-law approximations, and
we discuss how these differences impact both quintessence and Hubble
tension applications of the ULA models. Our main results are summarized in Sec. III.

\section{Evolution of ULA scalar fields}
The equation of motion for a scalar field with potential
$V(\phi)$ is given by

\begin{equation}
\label{phievol}
\ddot{\phi} + 3H\dot{\phi} + \frac{dV}{d{\phi}} = 0,
\end{equation}
where dot denotes the time derivative, and
the Hubble parameter $H$ is given by
\begin{equation}
    H \equiv \frac{\dot{a}}{a} = \sqrt{\rho_T/3}.
\end{equation}
In this equation, $a$ is the scale factor, $\rho_T$ is the total density, and we take $8 \pi G = c = \hbar = 1$ throughout.
The pressure and density of $\phi$ are given by
\begin{equation}
    p_\phi = \frac{\dot{\phi}^2}{2} - V(\phi),
\end{equation}
and 
\begin{equation}
    \rho_\phi = \frac{\dot{\phi}^2}{2} + V(\phi), 
\end{equation}
and
the equation of state parameter $w$ is defined as
\begin{equation}
    w = \frac{p_\phi}{\rho_\phi}.
\end{equation}

Now consider a scalar field $\phi$ evolving
in a potential of the form Eq. (\ref{ula}) or (\ref{powerula}).  We will assume that any nonzero value of $\dot \phi$ is damped by Hubble friction,
so that the scalar field
is initially at rest ($\dot \phi \approx 0$, $w \approx -1$).
As the field begins to thaw, it will
start to roll downhill.  At first,
the field energy will be dominated
by the potential, so that $\dot\phi^2/2 \ll V(\phi)$, but as $\dot \phi$ increases and $V(\phi)$ decreases, the field will eventually reach a state for which $V(\phi) \sim \dot \phi^2/2$.  Finally, after the field reaches the bottom of the potential, it will undergo rapid oscillations, with frequency $\nu \gg H$. 

While there is no general analytic solution for $\phi(t)$ in any of these regimes, there are well-known approximations for the two limiting cases:  the initial slow-roll regime with
$\dot \phi^2/2 \ll V(\phi)$, and the final
oscillatory phase with $\nu \gg H$.  These two cases are the subject of the next two subsections.

\subsection{Slow rolling ULA fields}

Here we will examine the initial ``slow-rolling" phase of a field evolving in a ULA potential.
As the field rolls downhill from its initial value in the potential $V(\phi)$, $w$ slowly increases, but $\dot \phi^2/2 \ll V(\phi)$, so that $w$
remains close to $-1$.
The evolution of $\phi$ in this slow-rolling regime depends on the relative values of $V$, $V^\prime$ and
$V^{\prime \prime}$,
where the prime indicates throughout derivative with respect to the scalar field, $\phi$.
Such fields can provide a natural mechanism to yield dark energy at late times with an equation of state $w \approx -1$.  In the terminology of Ref. \cite{CL}, these are ``thawing" quintessence fields.  If $\dot \phi^2/2$ never
becomes large compared to $V(\phi)$, then the field
evolves from $w = -1$ to a value of $w$ only slightly greater than $-1$, consistent with current observations.

Assuming that $w$ never diverges far
from $-1$, Ref. \cite{ScherrerSen} considered potentials satisfying the inflationary slow-roll
conditions, namely
\begin{equation}
\label{slow1}
\left(\frac{V^\prime}{V}\right)^2 \ll 1,
\end{equation}
and
\begin{equation}
\label{slow2}
\frac{V^{\prime\prime}}{V} \ll 1,
\end{equation}
while in Refs. \cite{ds1,Chiba,ds2}, the condition on the potential given by Eq. (\ref{slow1}) was retained,
but condition (\ref{slow2}) was relaxed.

When conditions (\ref{slow1}) and (\ref{slow2}) both are imposed on the potential, along with the thawing
initial condition ($\dot \phi \approx 0$ at early times), it is possible to derive an approximate analytic
solution for $w(a)$ that is independent of $V(\phi)$.  This solution is \cite{ScherrerSen}
\begin{equation}
\label{wlinear}
1+w(a) = (1+w_0)\frac{\left[F(a) - (F(a)^2-1) \coth^{-1} F(a) \right]^2}
{\left[ F(1) - (F(1)^2-1) \coth^{-1} F(1)  \right]^2},
\end{equation}
where $w_0$ is the value of $w$ at the present.
The function $F(a)$ is
\begin{equation}
F(a) = \sqrt{1+(\Omega_{\phi 0}^{-1} - 1)a^{-3}},
\end{equation}
where $\Omega_{\phi 0}$ is the fraction of the total density at present contributed by the scalar field.  With these
definitions, $F(a) = 1/\sqrt{\Omega_\phi(a)}$ and $F(1) = 1/\sqrt{\Omega_{\phi 0}}$.
Here and throughout we will not give detailed derivations of previously-derived results but will instead
cite the original papers; in this case,
a detailed derivation of Eq. (\ref{wlinear})
is given in Ref. \cite{ScherrerSen}.  The
evolution of $w(a)$ given by Eq. (\ref{wlinear}) is
well-approximated by a roughly linear dependence
of $w(a)$ on $a$:  the Chevallier-Polarski-Linder \cite{CP,L}
parametrization.  Specifically, we have
\begin{equation}
w(a) = w_0 +w_a(1-a),
\end{equation}
where the value of $w_a$ is well-fit by \cite{ScherrerSen}
\begin{equation}
w_a \approx -1.5(1+w_0)
\end{equation}

In Refs. \cite{ds1,Chiba,ds2}, the condition on the potential given by Eq. (\ref{slow1}) was retained,
but condition (\ref{slow2}) was relaxed, resulting in a wider range of possible behaviors.  In this
case, the evolution of $w$ with scale factor is given by \cite{ds1,Chiba,ds2}
	\begin{equation}
	\label{wquad}
	1 + w(a) = (1+w_0)a^{3(K-1)}\frac{[(F(a)+1)^K(K-F(a))
		+(F(a)-1)^K(K+F(a))]^2}
	{[(F(1)+1)^K(K-F(1))
		+(F(1)-1)^K (K+F(1))]^{2}},
	\end{equation}
where the constant $K$ is a function of $V^{\prime \prime}/V$ evaluated at $\phi_i$ (the
initial value of $\phi$), namely,
\begin{equation}
\label{Kdef}
K = \sqrt{1 - (4/3)V^{\prime \prime}(\phi_i)/V(\phi_i)}.
\end{equation}
Now instead of a single functional form for $w(a)$ for a given value of $w_0$, Eq. (\ref{wquad})
provides a family of solutions that depend on $K$.  As $K$ becomes large, these solutions
thaw more slowly, i.e., $w$ remains close to $-1$ until later in the evolution \cite{ds1}.
In the opposite limit, as $K \rightarrow 1$, the solution in Eq. (\ref{wquad}) approaches
the evolution given in Eq. (\ref{wlinear}). A graphical
representation of the trajectories for $w(a)$ for various
values of $K$ can be found in Refs. \cite{ds1,Chiba}.

With these results, we can examine the evolution of $w$ for ULA quintessence.  From Eq. (\ref{ula}) we derive
\begin{equation}
\label{deriv1}
\left(\frac{V^\prime}{V}\right)^2 = \left(\frac{n}{f}\right)^2 \frac{1+\cos(\phi/f)}{1-\cos(\phi/f)},
\end{equation}
and
\begin{equation}
\label{deriv2}
\frac{V^{\prime\prime}}{V} = \left(\frac{n}{f}\right)^2 \frac{1+\cos(\phi/f)-1/n}{1-\cos(\phi/f)},
\end{equation}
We see that $(V^\prime/V)^2 \ll 1$ is satisfied for two
cases:  either $f \gg n$, or $\phi_i/f \approx \pi$.
Now consider the value of $V^{\prime \prime}/V$ for
these two cases.
If $f \gg n$, then we have
${V^{\prime\prime}}/{V} \ll 1$.  This is the model
examined in Ref. \cite{ScherrerSen}, and it produces a single form for the evolution of $w(a)$, which is given in Eq. (\ref{wlinear}).  On the other hand, if $\phi_i/f \approx \pi$, then Eq. (\ref{deriv2}) gives
$V^{\prime \prime}/V = -(n/2f^2)$.
In this case $V^{\prime \prime}/V$ is necessarily negative, but it can be arbitrarily large or small, depending on the values of $n$ and $f$.  The corresponding value
of $K$ is
\begin{equation}                                 
K = \sqrt{1 + \frac{2}{3}\frac{n}{f^2}}.
\end{equation}
In this case, we can obtain the full range of possible
trajectories for $w(a)$ given in Eq. (\ref{wquad}), with $w(a)$ dependent
on the value of $n/f^2$.

Contrast this to the behavior of the corresponding power-law potentials in Eq. (\ref{powerula}).
For these potentials, we have
\begin{equation}
\left(\frac{V^{\prime}}{V}\right)^2 = \frac{4n^2}{\phi^2}
\end{equation}
and
\begin{equation}
\frac{V^{\prime\prime}}{V} = \frac{2n(2n-1)}{\phi^2}
\end{equation}
For these potentials, $(V^\prime/V)^2 \ll 1$ when $\phi_i \gg n$; when this is the case, $V^{\prime\prime}/V \ll 1$ as well.
Thus, the set of possible slow-roll quintessence evolutions is much more restricted than is the case for the ULA potentials.  For  slow-roll quintessence from power-law potentials, $w$ always evolves as in Eq. (\ref{wlinear}), and never as in Eq. (\ref{wquad}).  Recall that these power-law potentials are the limiting case of the ULA potentials when $\phi \ll f$.  This limiting case is consistent with slow-roll behavior when $n \ll \phi_i \ll f$.
                                                      
For the case in which the ULA field acts as early dark energy, the results in \cite{ScherrerSen,ds1,Chiba,ds2} must be
generalized to the case of a background (matter or
radiation) dominated expansion.  While this is a
straightforward calculation, the results are of little utility
in describing the resulting evolution of the density.
The reason is that in the early dark energy models
\cite{Poulin1,Poulin2,Agrawal,Lin,Smith,Murgia}, the regime
of interest occurs after the scalar field has
exited the slow-rolling regime and reached its maximum
density relative to background density.
If this occurs in the radiation-dominated era, then
this maximum density relative to the background is achieved
when $w = 1/3$, which is well beyond the validity of the
slow-roll approach.
However,
the subsequent oscillation and decline of the scalar
field energy density can be usefully described analytically, as shown in the next subsection.
	
\subsection{Oscillating ULA fields}

Rapidly oscillating scalar fields, in which the period of the scalar field oscillation is much less than
the Hubble time, were first systematically explored by Turner \cite{Turner}, and subsequently by many others
\cite{Kodama,liddle,sahni,hsu,masso,JK,dutta}.
Consider a rapidly-oscillating scalar field with potential $V(\phi)$.  We will consider only potentials
symmetric about $\phi=0$, so that the field oscillates between $\phi = -\phi_m$ and $\phi = +\phi_m$.
Then the energy density of this field is equal to $\rho_\phi = V(\phi_m)$.  Although $\phi_m$ and
 $\rho_\phi$ change with time, we will assume that this evolution is much slower than the oscillation frequency. Then the
equation of state can be expressed as \cite{Turner}
\begin{equation}
\label{wosc}
1+w = 2 \frac{\int_{0}^{\phi_m}[1-V(\phi)/V(\phi_m)]^{1/2}d\phi}{\int_{0}^{\phi_m}[1-V(\phi)/V(\phi_m)]^{-1/2}d\phi}.
\end{equation}
For power law potentials of the form given in Eq. (\ref{powerula}),
we obtain the standard result \cite{Turner}
\begin{equation}
\label{wpower}
w = \frac{n-1}{n+1}.
\end{equation}
Now consider the ULA potential (Eq. \ref{ula}).  To simplify our
expressions, we make the change of variables
$\Theta = \phi/f$ and obtain
\begin{equation}
\label{Turnerw}
1+w = 2 \frac{\int_{0}^{\Theta_m}[1-(1-\cos{\Theta})^n/(1-\cos{\Theta_m})^n]^{1/2}d\Theta}
{\int_{0}^{\Theta_m}[1-(1-\cos{\Theta})^n/(1-\cos{\Theta_m})^n]^{-1/2}d\Theta}.
\end{equation}
For $n=1$, this result for $1+w$ can be expressed in terms of elliptic integrals, but that provides
little insight into the behavior of the equation of state parameter.  Instead, we have numerically integrated Eq. (\ref{Turnerw}) for $n=1-3$; the results for $1+w$
are shown in Figs. $1-3$, respectively.

\begin{figure}
\includegraphics[width=.75\textwidth]{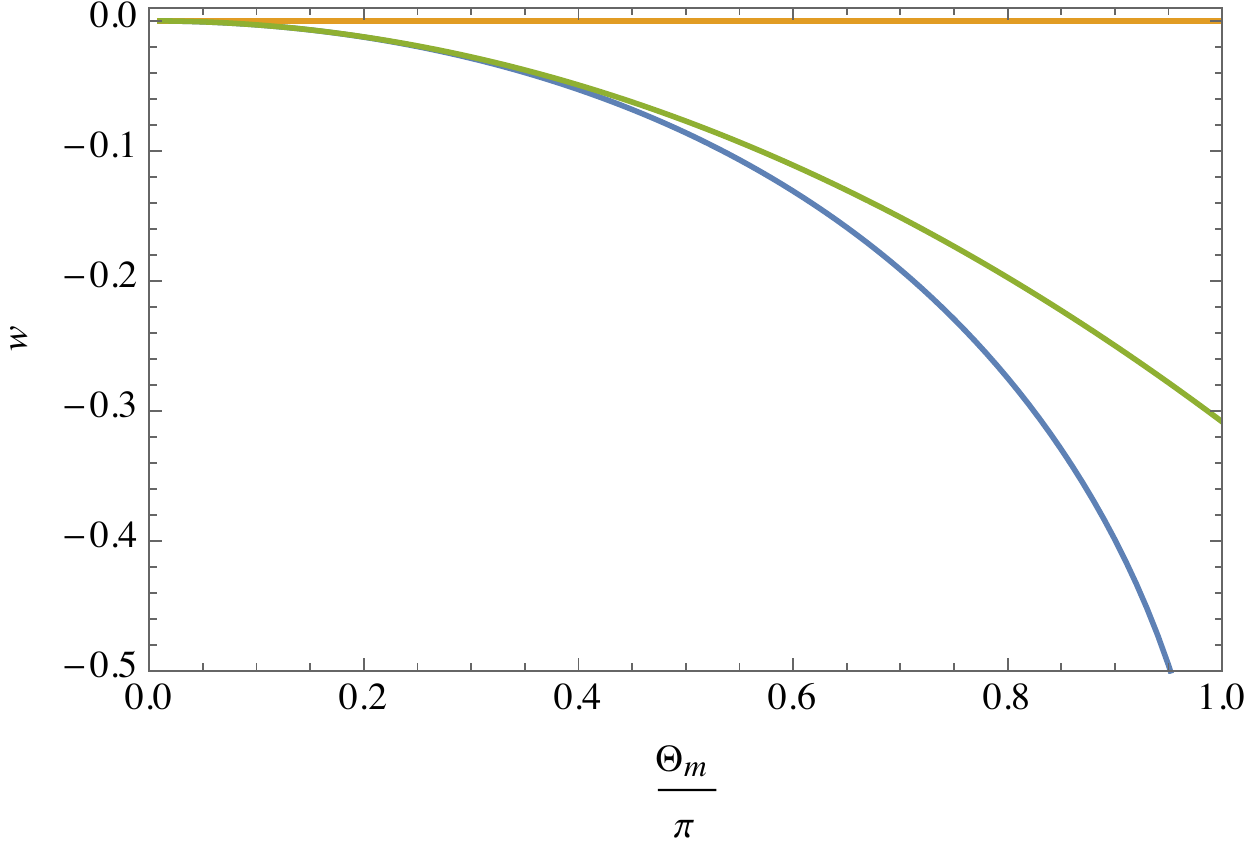}
\caption{The equation of state parameter $w$ as a function of $\Theta_m$, the maximum value of the displacement angle,
for the ULA potential (Eq. \ref{ula}) with $n=1$ (blue curve).  Gold curve gives the equation of state parameter for the power law potential that
corresponds to the ULA potential in the limit of small $\Theta_m$.  Green curve is the approximate ULA equation of state parameter to quadratic order in $\Theta_m$.}   
\end{figure}
\begin{figure}
\includegraphics[width=.75\textwidth]{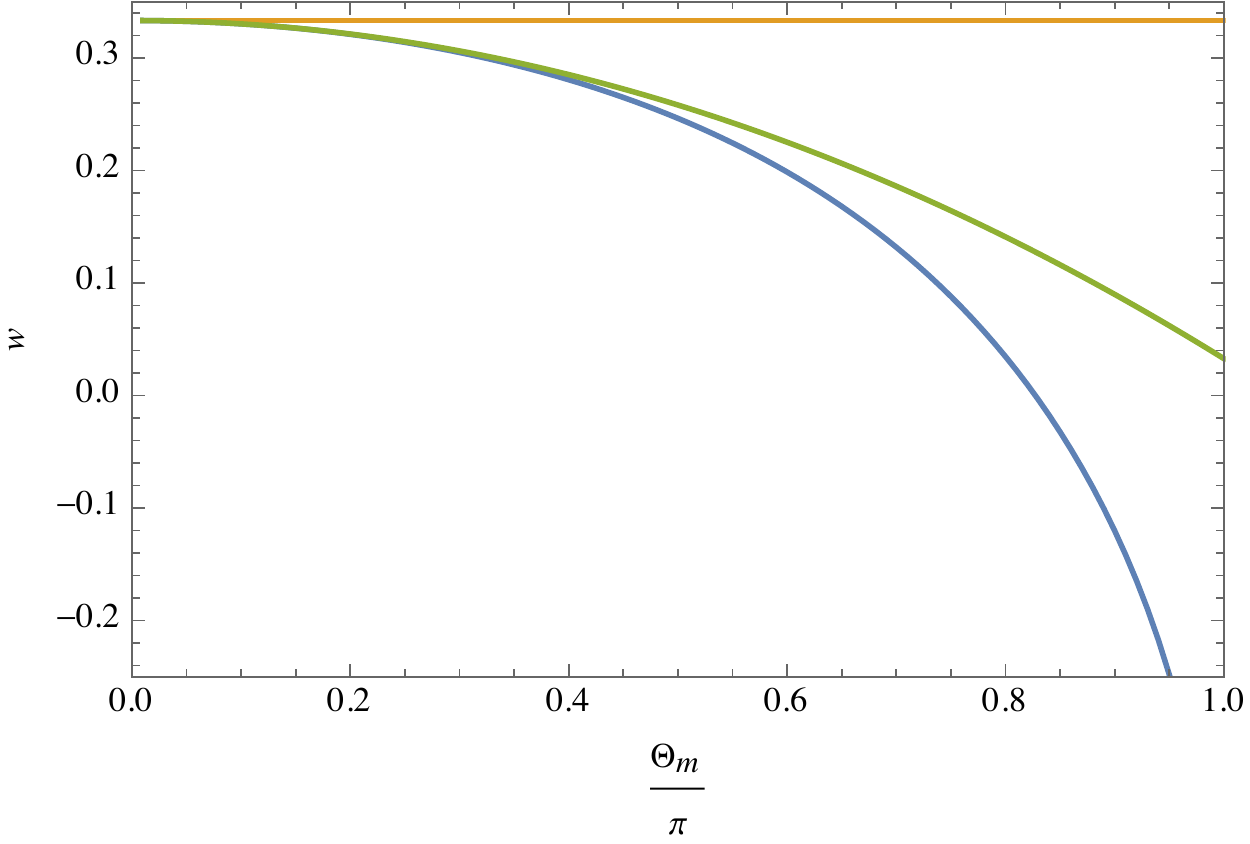}
\caption{As Fig. 1, for $n=2$.}    
\end{figure}
\begin{figure}
\includegraphics[width=.75\textwidth]{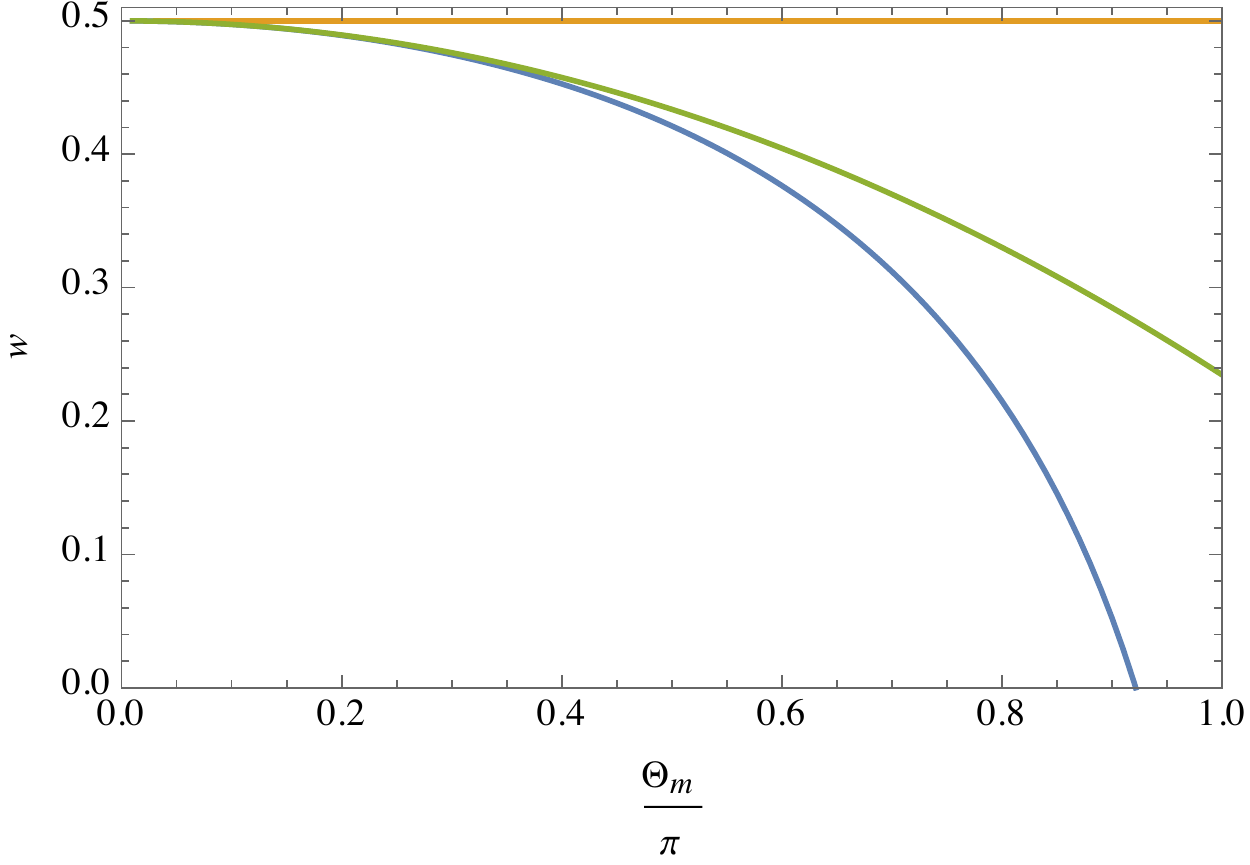}
\caption{As Fig. 1, for $n=3$.}    
\end{figure}

As expected, the equation of state parameters for the power law and ULA potentials are identical at small $\Theta_m$ and diverge at
large $\Theta_m$.  We can derive an analytic
approximation for $1+w$ for the ULA potentials at small
$\Theta_m$ using the expression
in Ref. \cite{Turner} for potentials that diverge slightly from power-law behavior.  Specifically, for potentials of the form
\begin{equation}
V(\phi)=a\phi^k(1+\epsilon \phi^l),
\end{equation}
with $\epsilon \ll 1$,
the equation of state parameter is given by \cite{Turner}
\begin{equation}
1+w = \frac{2k}{k+2} + \epsilon \phi_m^l\frac{4l(l+1)}{(2l + k + 2)(k+2)} \frac{\Gamma\left(\frac{k+2}{2k}\right)\Gamma\left(\frac{l+1}{k}\right)}{\Gamma\left(\frac{1}{k}\right)\Gamma\left(\frac{2l + k + 2}{2k}\right)},		
\end{equation}
plus higher-order terms. 

Expanding the ULA potential (Eq. \ref{ula}) to the lowest order beyond pure power-law behavior gives
\begin{equation}
V(\Theta) = m^2 f^2 \frac{\Theta^{2n}}{2^n}\left[1 - \frac{n}{12}\Theta^2\right],
\end{equation}
from which we can derive the corresponding approximations to
the equation of state parameter for small $\Theta_m$, namely, 
\begin{eqnarray}
	w &\approx& 0-\frac{1}{32}\Theta_m^2,~~~~~~~~~~~n=1,\\
    w &\approx& \frac{1}{3} - 0.0305 ~\Theta_m^2,~~~~~n=2,\\
    w &\approx& \frac{1}{2} - 0.0269 ~\Theta_m^2,~~~~~n=3.
\end{eqnarray}
These expressions are displayed in Figs. $1-3$.  Note
that the power-law value for $w$ is a poor approximation in all of
these cases once $\Theta/\pi$ increases beyond $\sim 0.1$.  The leading-order perturbation does much better, giving very good agreement
out to roughly $\Theta = \pi/2$.

Now consider the behavior of the oscillation frequency $\nu$.  For a potential $V(\phi)$, this frequency is given by \cite{masso,dutta}
\begin{equation}
\label{nu}
\nu = \left[\int_0^{\phi_m} 2 \sqrt{\frac{2}{V(\phi_m)-V(\phi)}}d\phi\right]^{-1}
\end{equation}
For the power-law potential in Eq. (\ref{powerula}), this expression can be integrated exactly to give \cite{masso}
\begin{equation}
\nu = \frac{n \Gamma(\frac{1}{2} + \frac{1}{2n})}{\sqrt{\pi} \Gamma(\frac{1}{2n})2^{(n+1)/2}} m\Theta_m^{n-1}.
\end{equation}
(See also the corresponding expression in Ref. \cite{Smith}).
The oscillation frequency for the ULA
potential can be derived numerically by substituting the potential from Eq. (\ref{ula}) into the integral for the frequency given by Eq. (\ref{nu}).  Figs. $4-6$ provide a comparison of the oscillation frequency for the ULA potential with the oscillation frequency for the corresponding small-$\Theta_m$
power-law potential.
\begin{figure}
	\includegraphics[width=.75\textwidth]{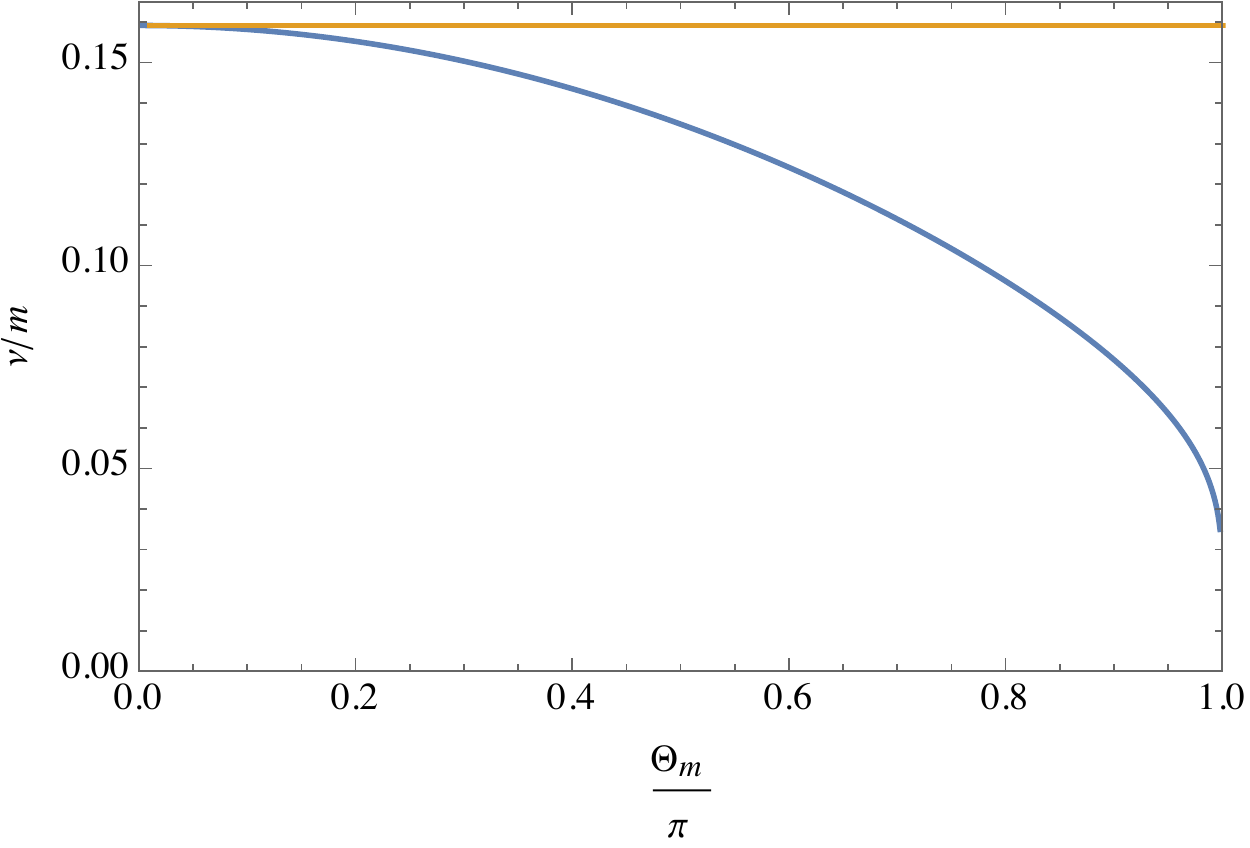}
	\caption{Oscillation frequency $\nu$ as a function of $\Theta_m$, the maximum value of the displacement angle,
		for the ULA potential (Eq. \ref{ula}) with $n=1$ (blue curve).  Gold curve gives the oscillation frequency for the power law potential that
	corresponds to the ULA potential in the limit of small $\Theta_m$.}   
\end{figure}
\begin{figure}
	\includegraphics[width=.75\textwidth]{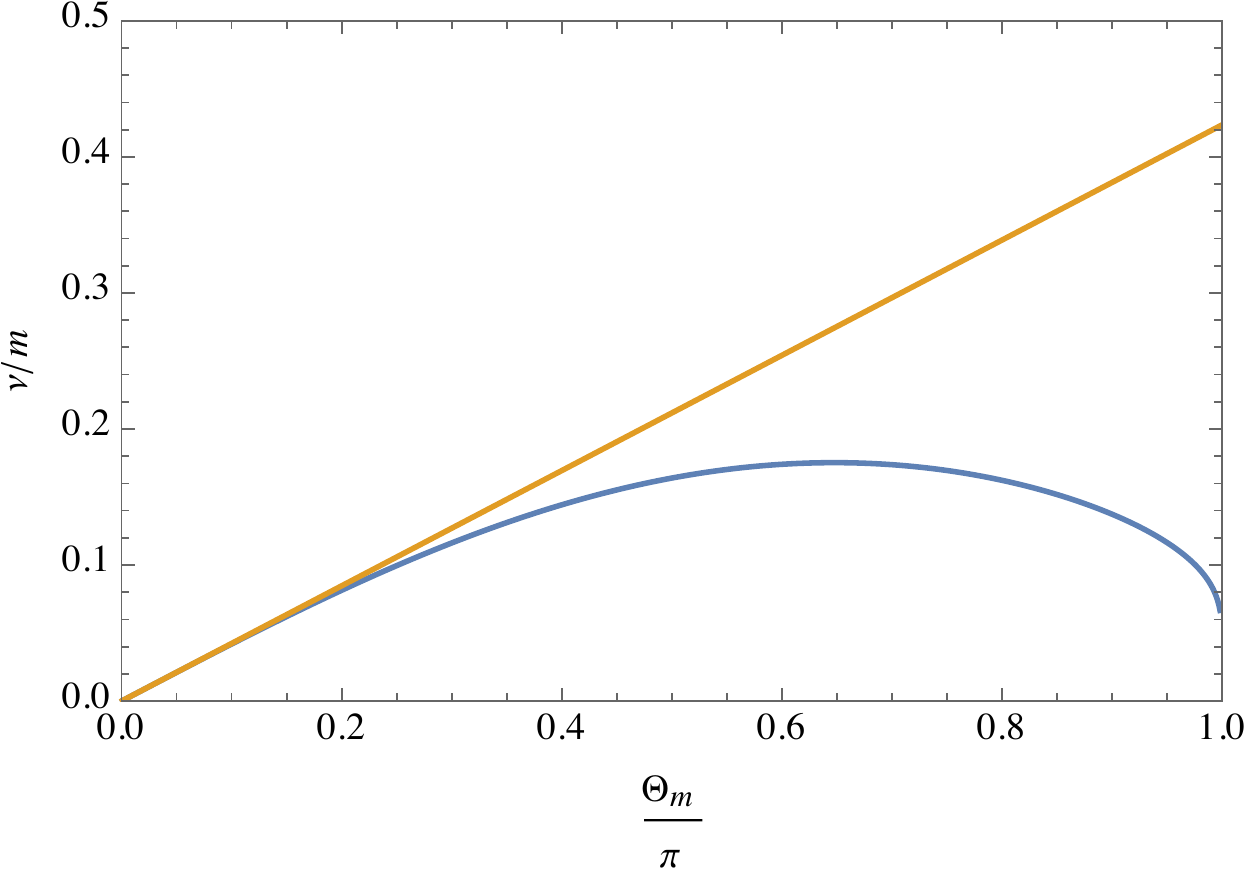}
	\caption{As Fig. 4, for $n=2$.}    
\end{figure}
\begin{figure}
	\includegraphics[width=.75\textwidth]{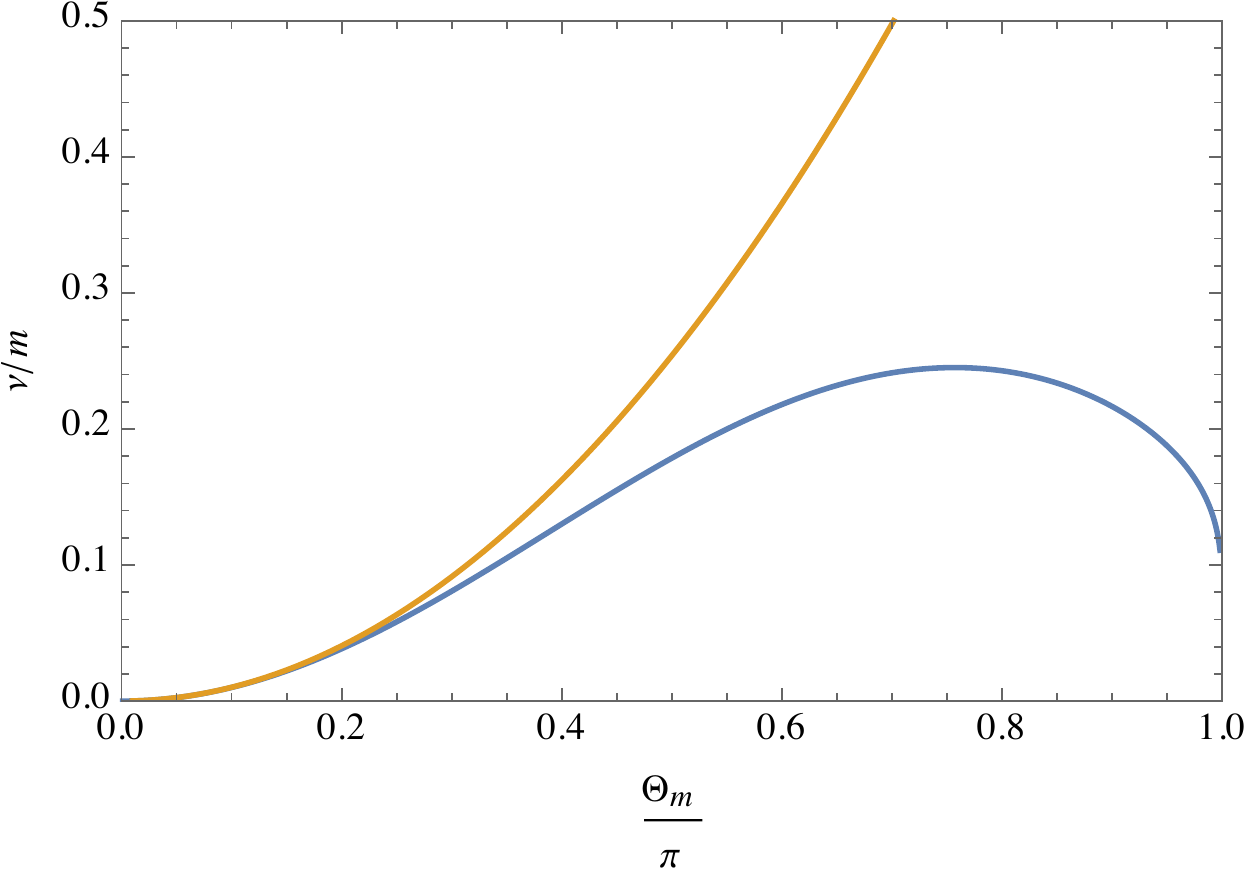}
	\caption{As Fig. 4, for $n=3$.}    
\end{figure}
As expected, the power law gives a good approximation for the oscillation frequency only for small values of $\Theta$, roughly
$\Theta_m/\pi \la 0.2$.

Now consider the cosmological consequences of these results.  For the ULA potentials with a rapidly-oscillating scalar field, the time-averaged value of $w$ approaches $-1$ as $\Theta_m$ goes to $\pi$.  In theory, such models could serve as dark energy (see, e.g., Ref. \cite{masso}).  However, Johnson and Kamionkowski \cite{JK} have argued that such models are unstable to growth of inhomogeneities.  Thus, quintessence models based on ULA potentials with high-frequency
oscillations are not as plausible
as the slow-roll quintessence models discussed in the previous section.

On the other hand, these rapidly oscillating ULA scalar fields form an important component of the early dark energy models
that resolve the Hubble tension
\cite{Poulin1,Poulin2,Agrawal,Lin,Smith,Murgia}.  Furthermore,
Smith et al. \cite{Smith} argue that the ULA potentials with
$n = 2-3$ (with $n=3$ providing the best fit) and large values
of $\Theta_m$ (specifically $\Theta_m \ga 2.5$) provide a better fit to the Hubble data than
the power-law potentials corresponding to $\Theta_m \ll 1$.

Our results
provide some insight into the evolutionary behavior of
the models examined in Ref. \cite{Smith}.  In particular,
the fact that $w$ decreases with increasing $\Theta_m$
for the ULA potentials
results in a slower decrease of the oscillation-averaged
density at low redshift for large values of $\Theta_m$.  This is clearly the case
in the numerical simulations (Fig. 2) of Ref. \cite{Smith}.
These simulations also suggest that the oscillation
frequency is much larger for the $n=2$ and $n=3$ ULA potentials with initial
values of $\Theta_m$ near $\pi$ than for very small initial values of $\Theta_m$.  This behavior is also apparent in our results (Figs. 5 and 6).  Further, for $n=2-3$,
large initial values of $\Theta_m$ yield an
oscillation frequency that increases as the scalar
field energy density (and therefore $\Theta_m$) decreases,
which is the opposite of the evolution of $\nu$ for very
small initial values of $\Theta_m$.

\section{Discussion}

Our analysis of the scalar field evolution for the full ULA potentials (Eq. \ref{ula}) shows many interesting differences from
the scalar field evolution for the corresponding small-$\phi$ power-law approximations (Eq. \ref{powerula}).  While
both sets of models can serve as slow-roll thawing quintessence, only the full ULA potentials can yield
hilltop-style evolution of $w$, corresponding to a much richer set of evolutionary behaviors.

For rapidly-oscillating
scalar fields, the
oscillation-averaged value for the equation of state parameter $w$ corresponding to the
power-law approximation diverges from the value
of $w$ in the ULA potential
for $\phi/f\pi > 0.1$, with
the ULA potential giving a value for $w$ much smaller than
the corresponding power-law potential, and
$w \rightarrow -1$ as $\phi/f \rightarrow \pi$.
Similarly, the power-law approximation for the
oscillation frequency $\nu$ diverges from the
ULA frequency for $\phi/f\pi > 0.2$,
with the ULA potential yielding a smaller oscillation
frequency.
Further, the dependence of the oscillation frequency
on the oscillation amplitude $\phi_m$ is more complex for
the ULA potentials; for $n=2$ and $n=3$, the oscillation
frequency increases with $\phi_m$, reaches a maximum, and then decreases as $\phi_m/f \rightarrow \pi$.

We emphasize that the standard axion potential ($n=1$) has been investigated previously, as has the behavior of the $n > 1$ potentials in the limit where they are well-approximated by a power-law potential.  What is new here is the treatment of the latter cases with the full ULA potential.  Furthermore, our quintessence modeling in Sec. II.A. is only approximate; it would be interesting to
investigate the behavior of these models in
the slow-roll regime with a full numerical integration of the equations of motion to more precisely determine their suitability as models for dark energy.

\begin{acknowledgments}

R.J.S. was supported in part by the Department of Energy (DE-SC0019207).  We thank Tristan Smith for helpful discussions.

\end{acknowledgments}


\begin{thebibliography}{99}


\bibitem{Lyth}
D.H. Lyth and A.A. Riotto, Phys. Rept. {\bf 314}, 1 (1999).

\bibitem{Allahverdi}
R. Allahverdi, R. Brandenberger, F.-Y. Cyr-Racine,
and A. Mazumdar, Ann. Rev. Nucl. Part. Sci. {\bf 60}, 27 (2010).

\bibitem{RatraPeebles}
  B. Ratra and P.J.E. Peebles,
  Phys.\ Rev.\  D {\bf 37}, 3406 (1988).
  
\bibitem{Wetterich}
C. Wetterich, Astron. Astrophys. {\bf 301}, 321 (1995).
  
\bibitem{Ferreira} P.G. Ferreira and M. Joyce,
\prl {\bf 79}, 4740 (1997).

\bibitem{CLW} E.J. Copeland, A.R. Liddle, and D. Wands,
\prd{\bf 57}, 4686 (1998).  
  

\bibitem{CaldwellDaveSteinhardt}
  R.R. Caldwell, R. Dave and P. J. Steinhardt,
  Phys.\ Rev.\ Lett.\  {\bf 80}, 1582 (1998).
  

\bibitem{Liddle}
  A.R. Liddle and R.J. Scherrer,
  Phys.\ Rev.\  D {\bf 59}, 023509 (1999).
  
  
\bibitem{SteinhardtWangZlatev}
  P.J. Steinhardt, L.M. Wang and I. Zlatev,
  Phys.\ Rev.\  D {\bf 59}, 123504 (1999).
  
\bibitem{Copeland1}
E.J. Copeland, M. Sami, and S. Tsujikawa, Int. J. Mod. Phys. D
{\bf 15}, 1753 (2006).


\bibitem{Freedman}
W.L. Freedman, Nat. Astron. {\bf 1}, 0121 (2017).

\bibitem{Knox}
L. Knox and M. Millea, \prd {\bf 101}, 043533 (2020).

\bibitem{Poulin1}
V. Poulin, T.L. Smith, D. Grin, T. Karwal, and
M. Kamionkowski,
Phys. Rev. D {\bf 98}, 083525 (2018).

\bibitem{Poulin2}
V. Poulin, T.L. Smith, T. Karwal, and M. Kamionkowski,
Phys. Rev. Lett. {\bf 122}, 221301 (2019).

\bibitem{Agrawal}
P. Agrawal, F.-Y. Cyr-Racine, D. Pinner, and L. Randall,
arXiv:1904.01016.

\bibitem{Lin}
M.-X. Lin, G. Benevento, W. Hu, and M. Raveri, \prd {\bf 100}, 063542 (2019).

\bibitem{Smith}
T.L. Smith, V. Poulin, and M.A. Amin,
\prd {\bf 101}, 063523 (2020).

\bibitem{Murgia}
R. Murgia, G.F. Abellan, and V. Poulin,
arXiv:2009.10733. 

\bibitem{Frieman}
J. Frieman, C. Hill, A. Stebbins, and I. Waga,
\prl {\bf 75}, 2077 (1995).

\bibitem{Kappl}
R. Kappl, H.P. Nilles, and M.W. Winkler,
Phys. Lett. B {\bf 753}, 653 (2016).

\bibitem{Capparelli}
L.M. Capparelli, R.R. Caldwell, and A. Melchiorri,
arXiv:1909.04621.


\bibitem{CL}
R.R. Caldwell and E.V. Linder, \prl {\bf 95}, 141301 (2005).

\bibitem{ScherrerSen}
R.~J.~Scherrer and A.~A.~Sen,
Phys.\ Rev.\  D {\bf 77}, 083515 (2008)

\bibitem{ds1}
S.~Dutta and R.~J.~Scherrer,
\prd {\bf 78}, 123525 (2008).

\bibitem{Chiba}
T. Chiba, \prd {\bf 79}, 083517 (2009).


\bibitem{ds2}
S.~Dutta, E.~N.~Saridakis and R.~J.~Scherrer,
\prd {\bf 79}, 103005 (2009).

\bibitem{CP}
M. Chevallier and D. Polarski, Int. J. Mod. Phys. D {\bf 10},
213 (2001).

\bibitem{L}
E.V. Linder, Phys. Rev. Lett. {\bf 90}, 091301 (2003).

\bibitem{Turner} M. S. Turner, Phys. Rev. D {\bf28}, 1243 (1983).

\bibitem{Kodama}
H. Kodama and T. Hamazaki, Prog. Theor. Phys. {\bf 96}, 949 (1996).

\bibitem{liddle}
A.R. Liddle and R.J. Scherrer,
\prd {\bf 59}, 023509 (1998).

\bibitem{sahni}
V. Sahni and L. Wang, \prd {\bf 62}, 103517 (2000).

\bibitem{hsu}
S.D.H. Hsu, Phys. Lett. B {\bf 567}, 9 (2003).

\bibitem{masso}
E. Masso, F. Rota, and G. Zsembinszki,
\prd {\bf 72}, 084007 (2005). 

\bibitem{dutta}
S. Dutta and R.J. Scherrer, \prd {\bf 78}, 083512 (2008).

\bibitem{JK} M.C. Johnson and M. Kamionkowski, \prd {\bf 78}, 063010 (2008).

\end{thebibliography}
\end{document}